\documentclass[longauth]{aa} % for the long lists of affiliations 
\def\kms{km\,s$^{-1}$}
\def\cd{d$^{-1}$}

\def\vsini{v$\sin$i}
\newcommand{\FeII}{\ion{Fe}{ii}}
\newcommand{\TiII}{\ion{Ti}{ii}}
\usepackage{graphicx}
\usepackage{txfonts}
\begin{document}
   \title{The $\gamma$\,Dor CoRoT target HD\,49434\\
          I-Results from the ground-based campaign\thanks{Based on observations made with ESO Telescopes at the La Silla Observatory under  the ESO Large Programme: LP178.D-0361, and on data collected
at the Centro Astron\'omico Hispano Alem\'an (CAHA) at Calar Alto, operated jointly by the Max-Planck
Institut f\"ur Astronomie and the Instituto de Astrof\'{\i}sica de Andaluc\'{\i}a (CSIC). Also based on observations obtained at Observatorio de Sierra Nevada (Spain), at Observatorio Astron\'omico Nacional San Pedro M\'artir (Mexico), at the Piszk\'estet\~{o} Mountain Station of Konkoly Observatory (Hungary), at Observatoire de Haute Provence (France) and at  Mount John University Observatory (New Zealand).}}

   \author{K.~Uytterhoeven \inst{1,2} 
           \and
           P.~Mathias \inst{3} 
           \and 
           E.~Poretti \inst{1} 
           \and 
	   M.~Rainer \inst{1}
	   \and
           S.~Mart\'{\i}n-Ruiz \inst{4}
           \and 
           E.~Rodr\'{\i}guez \inst{4}
	   \and
           P.J.~Amado  \inst{4}
           \and
	   D.~Le Contel \inst{3}
	   \and
           S.~Jankov \inst{3}
	   \and
	   E.~Niemczura \inst{5,6}
	   \and 
	   K.R.~Pollard \inst{7}
	   \and
	   E.~Brunsden \inst{7}
	   \and
           M.~Papar\'o \inst{8}
	   \and
           V.~Costa \inst{4}
           \and
	   J.-C.~Valtier \inst{3}
	   \and
	   R.~Garrido \inst{4}
	   \and 
	   J.C.~Su\'arez \inst{4}
	   \and
	   P.M.~Kilmartin \inst{7}
	   \and
	   E.~Chapellier \inst{3}
	   \and
	   C.~Rodr\'{\i}guez-L\'opez \inst{4}
	   \and
	   A.J.~Marin \inst{4}
	   \and
	   F.J.~Aceituno \inst{4}
	   \and
	   V.~Casanova \inst{4}
	   \and
	   A.~Rolland \inst{4}
	   \and
	   I.~Olivares \inst{4}
          }

   \offprints{K. Uytterhoeven}

   \institute{INAF-Osservatorio Astronomico di Brera, 
              Via E. Bianchi 46, 23807 Merate, Italy\\
              \email{katrien@iac.es}
         \and
Instituto de Astrof\'{\i}sica de Canarias, Calle Via L\'actea s/n, 38205 La Laguna, Spain \and
UMR 6525 H. Fizeau, UNS, CNRS, OCA, Campus Valrose, 06108 Nice Cedex 2, 
France 
\and 
	 Instituto de Astrof\'{\i}sica de Andaluc\'{\i}a (CSIC), Apartado 3004, 18080 Granada, Spain 
\and
        Institute of Astronomy, KULeuven, Celestijnenlaan 200D, 3001 Leuven, Belgium 
\and 
       Astronomical Institute of the Wroclaw University, ul. Kopernika 11, 51-622 Wroclaw, Poland
\and
   Department of Physics and Astronomy, University of Canterbury, Private Bag 4800, Christchurch, New Zealand 
\and
         Konkoly Observatory, P.O. Box 67, 1525 Budapest, Hungary 
}

  \date{Received ; accepted}

% \abstract{}{}{}{}{} 
% 5 {} token are mandatory
 
  \abstract
   {We present the results of an extensive ground-based photometric and 
    spectroscopic campaign on the $\gamma$\,Dor 
   CoRoT target HD\,49434. This campaign was preparatory to the CoRoT satellite
  observations, which took place from October 2007 to March 2008.
   }
   {Whereas satellite data will be limited to the detection of  low-degree pulsation modes
    with poor  identification (no filters), ground-based data will provide  
    eventually the identification of additional modes and, through the
    spectroscopic data, detection of additional high-degree modes as well as an  estimate of the azimuthal number $m$. Our aim was to detect and 
    identify as many pulsation modes as possible from the ground-based 
    dataset of the $\gamma$\,Dor star HD\,49434, to anticipate the CoRoT results.
   }
   {We searched for frequencies in the multi-colour variations, the pixel-to-pixel variations across the line profiles and the moments variations of 
 a large dataset consisting 
   of both multi-colour photometric and 
    spectroscopic data from different observatories, using different frequency analysis methods. We performed  
    a tentative mode identification of the spectroscopic frequencies using the Moment Method and the Intensity Period Search Method. We also carried out an abundance analysis.
   }
   {The frequency analysis clearly shows the presence of  four
 frequencies in the 0.2--1.7 \cd\,interval, as well as six frequencies in the 5--12\,\cd\, domain. The low frequencies
 are typical for $\gamma$\,Dor variables while the high frequencies
 are common for $\delta$\,Sct pulsators. We propose the frequency 2.666\,\cd\, as a possible rotational frequency. All modes, for which an identification was possible, seem to be high-degree modes ($3 \leq \ell \leq 8$). We did not find evidence for a possible binary nature of HD\,49434. The element abundances we derived are consistent with the values obtained in previous analyses.
%    The low frequencies are present in both the photometric and spectroscopic
%    data, whereas the high frequencies are detected in the line-profile diagnostics
%    only.
   }
   {We classify the $\gamma$\,Dor star HD\,49434 
    as a hybrid pulsator, which pulsates simultaneously in $p$-- and 
    $g$--modes. This finding makes  HD\,49434
    an extremely interesting target for  asteroseismic modelling.
    }

   \titlerunning{The $\gamma$\,Dor COROT target HD\,49434}
   \maketitle
%
%________________________________________________________________
\section{Introduction}
\label{intro}

We currently are in a challenging era of asteroseismology. Observers
and theoreticians are preparing for the promising output of the CoRoT
satellite mission (Baglin et al. 2006), which has been launched
successfully in December 2006. The CoRoT science programme is
twofold. A first goal of the mission involves the detection of
extrasolar planets by using the transit method. The asteroseismic
goal, on the other hand, aims at probing the internal structures of
stars by extracting information from detected pulsation frequencies
(several tens, even hundreds, of frequencies, depending on the
pulsational class), and, in general, to understand better the process
of stellar evolution. To this end, a selection of stars of a wide
range of masses, with different evolutionary status, of different
spectral type and belonging to different pulsation classes are
monitored.

The CoRoT satellite mission goes hand in hand with a huge
observational effort from the ground. Not only have preparatory
observations been a key stone in the selection
process of suitable targets for asteroseismology
 (Poretti et al. 2003, 2005), also now the
satellite is operating from space, simultaneous ground-based
observations are very important and complementary.  Indeed, as CoRoT
provides high-precision time-series in white light, multi-colour
photometry provides colour information, which allows identification of
the degree $\ell$ by means of amplitude ratios and phase shifts.
High-resolution spectroscopy allows the detection of high-degree modes
and the identification of both the degree $\ell$ and the azimuthal
order $m$ by means of, for example, the Moment or Doppler Imaging methods. With
the goal of obtaining multi-colour and spectroscopic time-series of a
selection of $\delta$\,Sct, $\gamma$\,Dor, $\beta$ Cep and Be CoRoT
primary and secondary targets, an extended ground-based campaign,
involving both high-resolution spectrographs and multi-colour
photometric instruments, has been included in the scientific plan of
the CoRoT mission (Catala et al. 2006; Uytterhoeven \& Poretti 2007;
Uytterhoeven et al. 2008).

The $\gamma$\,Dor stars are a challenging class of pulsators for
several reasons. They are a fairly recently discovered pulsational
group (Kaye et al. 1999) and, even though the number of class members
is rapidly increasing (more than 130 candidates are known), the number
of well-studied cases based on extended time-series is
small. Obtaining a good phase coverage from the ground is an observationally
challenging task as the intrinsic variations show periods of the order
of a day (0.4--3 days). Moreover, only a few frequencies have been
detected so far in individual stars, and amplitudes are fairly small
(below 0.05 mag; 2 \kms). Nevertheless, the $\gamma$\,Dor stars are
promising targets for seismic studies as they show $g$--modes, which can
be used to probe the deep interior of the star. Over the last few
years progress has been made in the understanding of the pulsational
mechanism. Probably a flux modulation induced by the upper convective
layer is the driving source for the pulsations (Guzik et al. 2000;
Dupret et al. 2004; Grigahc\`ene 2004). However, some  details
need clarification, such as the the thickness and the depth of the
convection zone and the importance of diffusion. 

The $\gamma$\,Dor stars (Spectral Type F0V-F2V) occupy an interesting
region along the Main-Sequence in the Hertzsprung-Russell (HR-)
diagram, being enclosed by the classical instability strip and the
instability strip of the solar-like stars.  An interesting
investigation concerns the presence of hybrid stars, i.e. stars that
exhibit both $p$--mode and $g$--mode oscillations, in the overlap
region of the $\gamma$\,Dor and $\delta$\,Sct instability strips
(Dupret et al. 2005).  Observational evidence for self-excited hybrid
stars is scarce, and less than a handful genuine hybrid candidates
exist, including HD\,8801 (Henry \& Fekel 2005), HD114839 (King et
al. 2006) and BD+18\,4914 (Rowe et al. 2006).  The hybrid stars are of
particular interest as both the envelope and the deep interior of the
star can be probed through the study of $p$--modes and $g$--modes,
respectively.

\begin{figure}
\centering
\resizebox{1.0\linewidth}{!}{\rotatebox{-90}{\includegraphics{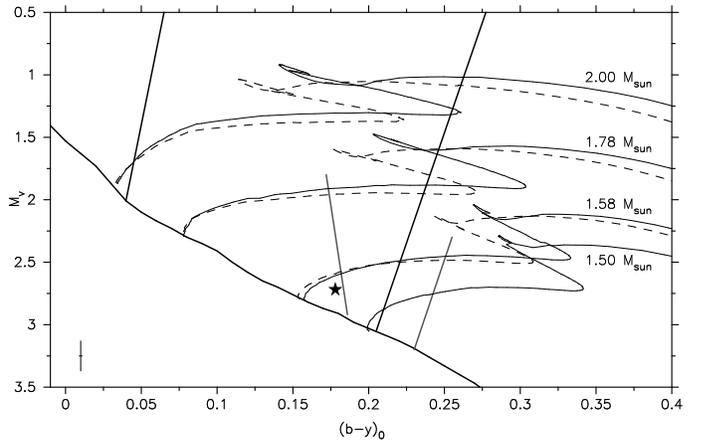}}}
\caption{Colour-Magnitude ($(b-y)_0$ versus Absolute Magnitude $M_v$) diagram indicating the position of HD\,49434 (star). The $(b-y)_0$ and $M_v$ values of HD\,49434  (an error bar is given in the left bottom corner) are taken from Hauck \& Mermilliod (1998) and from the HIPPARCOS satellite data, respectively. Dashed and solid tracks indicate evolutionary tracks for the overshooting extension distances $d_{over}$=0.1 and 0.2, respectively (see Claret 1995 for details). The ZAMS and the borders of the $\delta$\,Sct (longer black lines) and $\gamma$\,Dor (shorter gray lines) instability strips are indicated by solid lines.}
\label{CMD}
\end{figure}

In this paper we focus on the ground-based data of the $\gamma$ Dor
star HD\,49434 (V=5.75, F1V). As HD\,49434 is a primary CoRoT target
of a 150-days Long Run (LRa1; October 2007 -- March 2008), several
teams performed a detailed study of abundances and fundamental
parameters of this star in the preparatory framework of the
mission. Summarising their fairly consistent results, we find $T_{\rm
eff}=7300 \pm 200$ K, $\log g = 4.1 \pm 0.2$ dex, $[Fe/H] = -0.1 \pm
0.2$, \vsini $= 84 \pm 5$ \kms (Lastennet et al. 2001; Bruntt et
al. 2002, 2004; Mathias et al. 2004; Masana, Jordi \& Ribas
2006). Gillon \& Magain (2006) derived the slightly higher values
$T_{\rm eff} = 7632 \pm 126$ K and $\log g = 4.43 \pm 0.20$ dex.
In addition, we estimated these parameters from $uvby \beta$ and Geneva
photometry taken from the GCPD catalogue ({\it General Catalogue of
Photometric Data}, Mermilliod et al. 1997).  From the calibration of
Napiwotzki et al. (1993) we obtained $T_{\rm eff,[u-b]} = 7300$~K,
$T_{\rm eff,[b-y]} = 7230$~K and $\log g = 4.1$\,dex. From the Geneva
indices $(B2-V1)$, $d$ and $m2$ as proposed by K{\"u}nzli et
al. (1997), we determined $T_{\rm eff} = 7200 \pm 60$~K and $\log g =
4.34 \pm 0.08$\,dex.  Taking into account all these values of the
atmospheric parameters, we will adopt in our analysis 
effective temperature and surface gravity equal to $7300$~K and
$4.2$~dex, respectively. From 2MASS IR photometry Masana, Jordi \&
Ribas (2006) estimate an angular semi-diameter of $0.186 \pm 0.002$
mas, and $R = 1.601 \pm 0.052 R_{\odot}$. An estimate of the mass of
HD\,49434 $M = 1.55 \pm 0.14 M_{\odot}$ is obtained from evolutionary
tracks by Bruntt et al. (2004). A first study of the intrinsic
variability of HD\,49434 is presented by Bruntt et al. (2002). In a
time series of Str\"omgren data they found no clear frequencies, but
reported a power excess in the frequency range typical for
$\gamma$\,Dor pulsators. A line-profile analysis suggested the
presence of one or several high-degree modes, in the same low
frequency range. Line-profile variations were confirmed by Mathias et
al. (2004). In Fig.~\ref{CMD} we show the position of HD\,49434 in a
Colour-Magnitude diagram, based on the $M_v$ value derived from the
HIPPARCOS satellite and $(b-y)_0$, calculated from the photometric
indices taken from Hauck \& Mermilliod (1998) using the TempLogg
method (Kupka \& Bruntt 2001). We used the evolutionary tracks,
instability strips and ZAMS as described by Poretti et al. (2003). As
can be seen, HD\,49434 lies close to the blue border of the
$\gamma$\,Dor instability strip, and inside the instability strip of
$\delta$\,Sct stars.

\section{Observations}
\subsection{Multi-colour photometry}
 Str\"omgren observations were made with the twin Danish six-channel
 {\sl uvby}$\beta$ photometers at Sierra Nevada Observatory (SNO),
 Spain, and at San Pedro M\'{a}rtir Observatory (SPMO), Mexico. The
 instruments are mounted on the 90-cm and 1.5-m telescopes,
 respectively.  The data were collected in the four {\sl uvby} filters
 and exposure times were 30\,s. We used HD\,48922 ($V$=6.77\,mag,
 A0) as comparison star and HD\,43913 ($V$=7.88\,mag, A0) as check
 star. In total we obtained 1353 data points of HD\,49434 in 62 clear
 nights at SNO (January 2005; December 2005; January--March 2006;
 November 2006; January--March 2007; November--December 2007; January
 2008) and 614 data points in 29 clear nights at SPMO (November 2005;
 November--December 2006; November-December 2007). We note that an
 important step in the reduction process was the correction for the
 instantaneous extinction coefficient, in order to suppress
 artificial/spurious frequencies in the 0--5\,\cd\, range (Poretti \&
 Zerbi 1993).  The standard deviations of the magnitude differences
 between the comparison and check stars (both measured
 in the same observational cycle) give an indication of the precision
 of the data. We obtained values near 12.3\,mmag and 6.5\,mmag ($u$
 light), 3.0\,mmag and 4.2\,mmag ($v$ light), 3.0\,mmag and 3.8\,mmag
 ($b$ light), and 3.0\,mmag and 4.7\,mmag ($y$ light) for SPMO and
 SNO, respectively. The large scatter observed in the $u$--light at SPMO
 is due to some instrumental problems with the photo-multiplier of the $u$ channel.

\onltab{1}{
\begin{table*}
\caption[]{Journal of the photometric observations of HD\,49434 obtained between March 2003 and March 2007 at SNO, SPMO and KO. For each date (corresponding to UT at the beginning of the observations), the JD (-2450000) and the number of data points are given.}
\begin{center}
\begin{tabular}{lcccc|lcccc|lcccc}
\hline  \hline
Date & JD & \multicolumn{3}{c|}{$\#$ data points} & Date & JD & \multicolumn{3}{c|}{$\#$ data points} & Date & JD & \multicolumn{3}{c}{$\#$ data points} \\
     &    & {\tiny SNO} & {\tiny SPMO} & {\tiny KO}      &    &  & {\tiny SNO} & {\tiny SPMO} & {\tiny KO}      &    &  & {\tiny SNO} & {\tiny SPMO} & {\tiny KO} \\
\hline
21Mar2003& 2720.2 &  & & 4 	  & 03Mar2006& 3798.4 & 13  &   &   	  &  13Mar2007& 4173.3 & 9  &   &  \\       	  
22Mar2003& 2721.2 &  & & 7 	  & 06Mar2006& 3801.4 & 7   &   &  	  &  17Mar2007& 4177.3 & 16 &   &  \\
23Mar2003& 2722.2 &  & & 7 	  & 07Mar2006& 3802.3 & 17  &   &   	  &  18Mar2007& 4178.3 & 9  &   &  \\
	      24Mar2003& 2723.2 &  & & 7 	  & 08Mar2006& 3803.3 & 10  &   &  	  &  22Mar2007& 4182.3 & 15 &   &  \\
25Mar2003& 2724.2 &  & & 6 	  & 10Nov2006& 4049.5 & 30  &   & 	  &  15Nov2007& 5519.6 & 20 &   &  \\  	  
20Jan2005& 3391.3 & 32 & &     	  & 11Nov2006& 4050.5 & 25  &   &	  &  16Nov2007& 5520.5 & 25 &   &  \\ 
21Jan2005& 3392.3 & 57 & &     	  & 12Nov2006& 4051.5 & 32  &   &  	  &  17Nov2007& 5521.5 & 21 &   &  \\  	  
22Jan2005& 3393.3 & 59 & &     	  & 13Nov2006& 4052.5 & 32  &   &    	  &  28Nov2007& 5533.4 & 39 &   &  \\	      
12Nov2005& 3686.7 & & 16 &	  & 14Nov2006& 4053.5 & 36  &   &  	  &  30Nov2007& 5535.4 & 39 &   &  \\	    	      
13Nov2005& 3687.7 & &  22 &	  & 15Nov2006& 4054.6 & 17  &   &    	  &  02Dec2007& 5537.4 & 12 &   &  \\	  
14Nov2005& 3688.7 & & 25 &	  & 18Nov2006& 4057.6 & 15  &   &  	  &  03Dec2007& 4437.8 &    & 19&  \\	       	  
15Nov2005& 3689.7 & & 23 &	  & 20Nov2006& 4059.6 & 21  &23 &         &  04Dec2007& 4438.6 & 12 & 27 &  \\	  
06Dec2005& 3711.4 & 26 & &     	  & 21Nov2006& 4060.6 & 9   &17 &         &  05Dec2007& 4439.6 & 14 & 28 &  \\	    	      	
07Dec2005& 3712.4 & 31 & &     	  & 22Nov2006& 4061.8 &     & 18&         &  06Dec2007& 4440.6 & 12 & 22 &  \\	    	      	
12Dec2005& 3717.4 & 10 & &     	  & 23Nov2006& 4062.6 & 18  & 9 &         &  07Dec2007& 4441.9 &    &  7 &  \\	
13Dec2005& 3718.4 & 8  & &     	  & 24Nov2006& 4063.6 & 13 & 16  &    &      08Dec2007& 4442.8 &    &  5 &  \\
14Dec2005& 3719.4 & 10 & &     	  & 26Nov2006& 4065.7 &    & 15  &    &      10Dec2007& 4444.6 & 18 &    &  \\
15Dec2005& 3720.4 & 9  & &     	  & 27Nov2006& 4066.7 &    & 13  &    &      11Dec2007& 4445.6 & 14 &    &  \\
17Dec2005& 3722.4 & 11 & &     	  & 28Nov2006& 4067.7 &    &  22 &    &      12Dec2007& 4446.6 & 10 &    &  \\
18Dec2005& 3723.4 & 5  & &     	  & 29Nov2006& 4068.7 &    &  30 &    &      13Dec2007& 4447.7 &    & 25 &  \\
20Dec2005& 3725.4 & 26 & &    	  & 30Nov2006& 4069.7 &    &  20 &    &      14Dec2007& 4448.8 &   & 16 &  \\
21Dec2005& 3726.3 & 49 & &        & 01Dec2006& 4070.7 &    &  30 &    &      15Dec2007& 4449.8 &   & 16 &  \\
22Dec2005& 3727.4 & 20 & &     	  & 02Dec2006& 4071.7 &    &  31 &    &      29Dec2007& 4463.6 & 10&    &  \\
02Jan2006& 3738.4 & 11 & &	  & 03Dec2006& 4072.7 &    &  28 &    &      31Dec2007& 4465.6 & 12&    &  \\
04Jan2006& 3740.4 & 14 & & 	  & 04Dec2006& 4073.7 &    &  25 &    &      10Jan2008& 4476.4 & 33&    &  \\
11Jan2006& 3747.3 & 20 & &	  & 05Dec2006& 4074.7 &    &  25 &    &      12Jan2008& 4478.3 & 42&    &  \\
01Feb2006& 3768.3 & 33 & &	  & 06Dec2006& 4075.7 &    &  28 &    &      18Jan2008& 4484.4 & 35&    &  \\
04Feb2006& 3771.3 & 47 & & 	  & 24Jan2007& 4125.3 & 44 &     &    &      19Jan2008& 4485.3 & 37&    &  \\
27Feb2006& 3794.4 & 12 & &	  & 26Feb2007& 4158.3 & 15 &     &    &      20Jan2008& 4486.3 & 37&    &  \\
01Mar2006& 3796.4 & 15 & &	  & 11Mar2007& 4171.3 & 12 &     &    &               &        &   &    &  \\
\hline
\end{tabular}
\end{center}
\label{logbookphot}
\end{table*}}

In addition, Johnson UBVRI measurements were obtained with a single
channel Peltier-cooled photo-electric photometer on the 50-cm
Cassegrain telescope at the Piszk\'estet\~{o} Mountain Station of
Konkoly Observatory (KO) from March 21 to 25, 2003.  The comparison
star used was again HD~48922, as well as HD\,49933 ($V$=5.78\,mag,
F2V). The latter comparison star, a solar-like oscillator, is actually
a primary target of the CoRoT mission. Exposure times for individual
data points were 15\,s. Consecutive three data points were binned
together to create 31 measurements with 45\,s integration time
each.

\begin{figure*}
\centering
\resizebox{0.95\linewidth}{!}{\rotatebox{-90}{\includegraphics{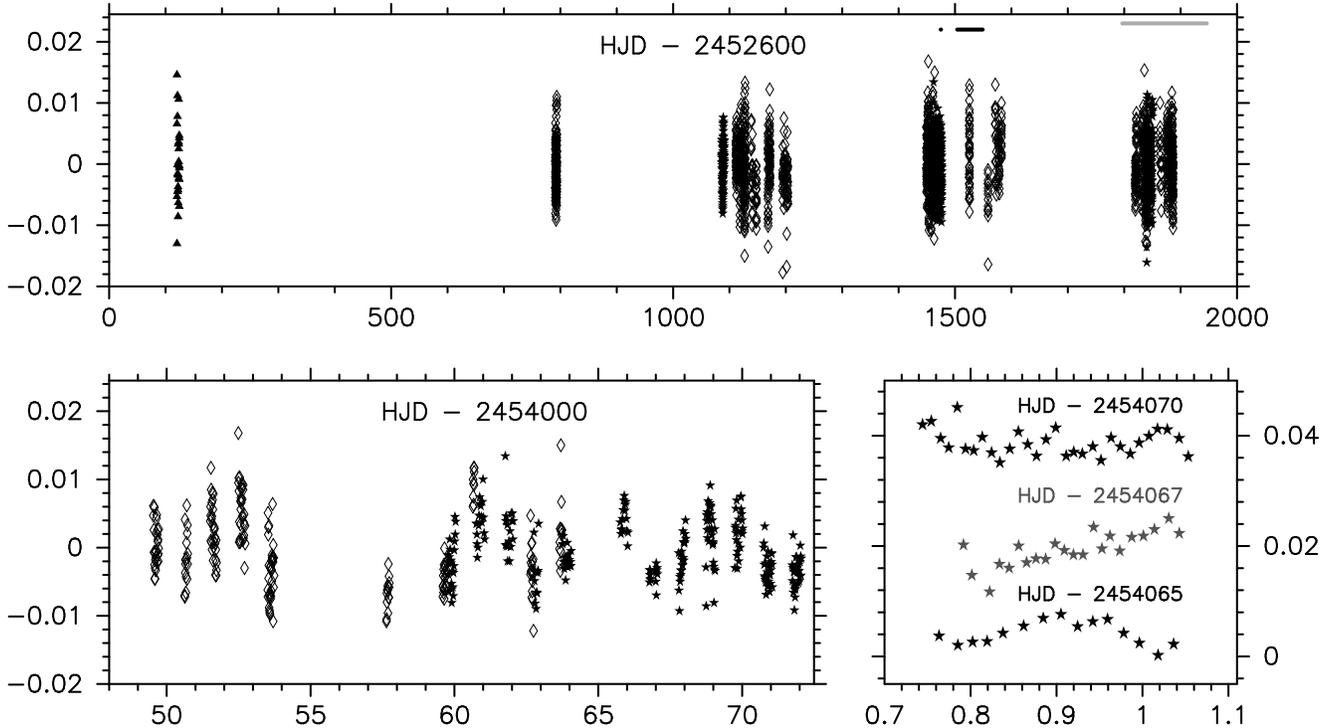}}}
\caption{Ground-based lightcurves of HD\,49434, showing variability at
different timescales. The top panel shows all relative KON Johnson V
(triangles), SNO Str\"omgren $y$ (open diamonds) and SPMO Str\"omgren
$y$ (stars) magnitudes. The bottom left panel zooms in on SNO and SPMO
data obtained in November 2006. The bottom right panel shows the
lightcurves obtained during the nights of 26 and 28 November 2006, and
1 December 2006 at SPMO. We plotted the individual nights with an
offset of 2.1 mmag for clarity. On top of the figure we
indicated the time span of the spectroscopic ground-based observations
described in this paper (see Sect.\,\ref{highres}) with a black line,
while the continuous CoRoT observations are given in gray.}
\label{lightcurves}
\end{figure*}

A logbook of the photometric observations is given in
Table~\ref{logbookphot}, which is only available in the on-line
version of the paper. An example of the lightcurve is given in
Fig.~\ref{lightcurves}. The differential lightcurves of HD\,49434 show
a complicated variable behaviour with variations at different
timescales.

We also included the data of the HIPPARCOS satellite (Perryman et
al. 1997) in our photometric analysis.

\subsection{High-resolution spectroscopy}
\label{highres}
The spectroscopic observations include data taken with FEROS@2.2-m
ESO/MPI telescope, La Silla, Chile, with SOPHIE@1.93-m telescope at
Observatoire de Haute Provence (OHP), France, with FOCES@2.2-m at
Calar Alto Astronomical Observatory (CAHA), Spain, and with
HERCULES@1.0-m McLellen Telescope at Mount John University Observatory
(MJUO), New Zealand.  A general logbook of the spectroscopic data of
HD\,49434 is given in Table~\ref{logbook}, while a log file of the
individual observing nights, only available in the on-line version of
the paper, is presented in Table\,\ref{log}. The time
distribution of the spectroscopic observations can be seen from
Fig.~\ref{moments}. The time-span with respect to the photometric
observations is indicated by a black full line in
Fig.~\ref{lightcurves}.

\begin{table*}
\caption[]{Logbook of the spectroscopic observations of HD\,49434
obtained in December 2006 -- February 2007 (HJD
2454074.4--2454149.1; $\Delta T = 75$ days). For each instrument the number of high-quality
spectra (i.e. spectra with a signal-to-noise (SN-) ratio $>$ 80 in
the regions near 4900 \AA\ and 5700 \AA), the number of successful nights, the time-span (days), the
average SN-ratio (SN), the range of SN-ratios (SN-range), the typical exposure
times (seconds), the resolution of the spectrograph, and the spectral range of the spectra (\AA) are given. The last column indicates if there are  gaps in the spectral coverage (y/n). }
\begin{center}
\begin{tabular}{cccccccccc}
\hline  \hline
 instrument & \# & \# nights  & $\Delta$T & SN & SN-range & Texp & resolution & range & spectral gaps\\ 
\hline
FOCES & 47 & 1 & $0^d.3$ & 185 & $[110,230]$ & 300& 40,000 & 3500--10280 & y, redder then 8450 \AA\\
FEROS & 71 & 14 & $26^d.1$ &  240 & $[100,320]$ & 120,180 & 48,000 & 3500--9200 & n\\
SOPHIE & 377 & 14 & $20^d.2$ & 165 & $[80,240]$ & 600,500,450& 70,000 & 3870--6940& n\\
HERCULES & 194 & 9 & $16^d.1$ &  172 & $[110,230]$ & 600,480,420& 35,000 & 4500--7300 & y\\
\hline
\end{tabular}
\end{center}
\label{logbook}
\end{table*}

\subsubsection{The SOPHIE instrument}
The new SOPHIE spectrograph, attached to the 1.93-m telescope at OHP,
is the successor of ELODIE.  One of its main properties is its
stability for performing asteroseismic observations.  The SOPHIE
spectrum, recorded on 39 non-overlapping spectral orders, ranges from
3870 to 6940 \AA.  The spectra were extracted and automatically reduced
in real-time using a data reduction software package directly adapted
from HARPS.  Only the normalisation process was conducted manually,
after a correction to the heliocentric frame, typically using a fit with a cubic spline.

Telescope time was attributed from January 11-20 and January 27-
February 1, 2007, representing 14 nights.
Only 1.5 night was lost due to meteorological conditions.
The 410 SOPHIE spectra of HD\,49434 were obtained in the High-Resolution
mode ($R\sim70\,000$), as we want to focus on the study of line-profile variations (LPV).

\onltab{3}{
\begin{table*}
\caption[]{Journal of the spectroscopic observations of HD\,49434 obtained between December 2006 and February 2007 with FEROS, SOPHIE, FOCES and HERCULES. For each date (corresponding to UT at the beginning of the observations), the JD (-2450000) and the number of spectra (with SN-ratio $>$ 80) are given.}
\begin{center}
\begin{tabular}{ccc|ccc|ccc}
\hline  \hline
\multicolumn{3}{c|}{FEROS} & \multicolumn{3}{c|}{SOPHIE} & \multicolumn{3}{c}{FOCES} \\ \hline
Date & JD & $\#$ spectra & Date & JD & $\#$ spectra & Date & JD & $\#$ spectra \\ \hline
03Jan2007 & 4103.7 & 5  & 11Jan2007 & 4112.4& 22  & 04Dec2006 & 4074.4 & 50 \\ 
04Jan2007 & 4104.6 & 7  & 12Jan2007 & 4113.4& 30  &           &        &    \\ 
05Jan2007 & 4105.6 & 6  & 13Jan2007 & 4114.4& 40  &            &        &    \\ \cline{7-9}
06Jan2007 & 4106.6 & 5  & 14Jan2007 & 4115.4& 30  & \multicolumn{3}{c}{HERCULES} \\ \cline{7-9}  
07Jan2007 & 4107.6 & 5  & 15Jan2007 & 4116.4& 18  & Date & JD & $\#$ spectra \\ \cline{7-9}  
08Jan2007 & 4108.6 & 6  & 17Jan2007 & 4118.4& 19  & 01Feb2007 & 4132.9 & 6 \\	       	    	
09Jan2007 & 4109.5 & 8  & 18Jan2007 & 4119.4& 13  & 04Feb2007 & 4135.9 & 19 \\	       	    	
10Jan2007 & 4110.5 & 7  & 19Jan2007 & 4120.4& 24  & 05Feb2007 & 4136.9 & 23 \\	       	    	
11Jan2007 & 4111.6 & 5  & 20Jan2007 & 4121.4& 23  & 06Feb2007 & 4137.9 & 26 \\	       	    	
25Jan2007 & 4125.6 & 3  & 27Jan2007 & 4128.3& 31  & 12Feb2007 & 4143.9 & 11 \\	       	    	
26Jan2007 & 4126.6 & 4  & 28Jan2007 & 4129.3& 36  & 13Feb2007 & 4144.9 & 22 \\	       	    	
27Jan2007 & 4127.6 & 5  & 29Jan2007 & 4130.3& 41  & 14Feb2007 & 4145.9 & 28 \\	       	    	
28Jan2007 & 4128.6 & 5  & 30Jan2007 & 4131.3& 42  & 16Feb2007 & 4147.9 & 30 \\	       	    	
29Jan2007 & 4129.7 & 2  & 31Jan2007 & 4132.3& 41  & 17Feb2007 & 4148.9 & 29 \\                  
\hline
\end{tabular}
\end{center}
\label{log}
\end{table*}}

\subsubsection{FEROS data}
\label{FEROSsec}
The Fiber-fed, Extended Range, \'Echelle  Optical Spectrograph (FEROS), mounted
at the 2.2-m ESO/MPI telescope at La Silla (ESO), has a resolution of $R
\sim 48,000$ and records almost the complete range of 3500--9200 \AA\ on
39 \'echelle orders. The detector is a EEV 2kx4k CCD. For the
observations of HD\,49434 we used the following set-up: the object+sky
fiber combination, 1x1 binning mode and the Atmospheric Dispersion
Corrector enabled. We reduced the spectra using an improved version of
the standard FEROS pipeline, written in MIDAS, developed by Rainer
(2003). The main improvements of this pipeline concern the blaze and
flat-field correction of the spectra, by using an accurate definition
of the blaze function extracted from a well-exposed spectrum of a hot
star. The reduced spectra were subsequently corrected to the
heliocentric frame.  We used an automated continuum normalisation
procedure developed by M.~Bossi (INAF OAB-Merate) to normalise the
spectra with continuum at unity. As the resulting normalised spectra
have to be treated with caution when performing a study of individual
line-profiles, selected spectral regions (e.g.\ near \TiII\ 4501.273
\AA\ and \FeII\ 4508.288 \AA) were also normalised manually by fitting
cubic splines.

HD\,49434 was one of the selected CoRoT targets observed during the
15 nights, from January 1--10 and from January 24--28, 2007,
awarded in the framework of an ESO Large Programme (LP178.D-0361; PI: E. Poretti).
We could observe 85\% of the total available time in
good weather conditions,  resulting in  73 spectra.

\subsubsection{HERCULES data}
The High Efficiency and Resolution Canterbury University Large Echelle
Spectrograph (HERCULES) is fibre-fed from the 1.0-m McLellan telescope
at MJUO. The instrument has a resolution of $R\sim35000$ and records
selected regions of the spectral range of 3800--8800 \AA\ on 48
orders. A description of the gaps in the spectral coverage is given by
Hearnshaw et al. (2002). Unfortunately, the \TiII\ and \FeII\ profiles
near 4500 \AA\ belong to one of those gaps.  The spectra were reduced
with a semi-automated procedure using the HERCULES Reduction Software
Package version 2.3 (Skuljan 2004). After a correction to the
heliocentric frame, the spectra were normalised using the automated
continuum normalisation procedure as described in Sect.~\ref{FEROSsec}.

Observing time was awarded on HERCULES from February 1--18, 2007, but
half of the nights were lost due to bad weather.  In total 194 spectra were observed. Exposure times ranged from 8 to 10 minutes in clear sky conditions, and up to 20
minutes when observing through thin clouds.  

\subsubsection{FOCES data}
The FOCES \'echelle spectrograph on the 2.2-m telescope at CAHA has a
spectral coverage of 3820--10280\AA\, in 93 spectral orders. The CCD is a
Loral 2kx2k CCD with a pixel size of 15$\mu$m.  For the observations
of the $\gamma$\,Dor target we used the $R\sim40,000$ mode.  As there
is no pipeline available, the spectral reduction and subsequent
heliocentric correction were performed using standard IRAF \'echelle
spectral reduction procedures.  The spectra were normalised using an 
automated continuum normalisation procedure.

A total of 25 nights were awarded on FOCES to observe HD\,49434. The
time was distributed over two campaigns in December 2006 and 2007,
with 5 more nights in February 2008. Unfortunately, bad weather
hampered these runs reducing the usable data to only one night. A
total of 47 FOCES spectra of HD\,49434 were observed on December 4,
2006. 

\section{Analysis}
\subsection{Multi-colour analysis}
\label{multi-colour}
The Str\"omgren dataset consists of 4 seasons of SNO observations and
3 seasons of SPMO data, of which the most continuous dataset (361
datapoints in 18 nights) was observed at SPMO in November-December
2006 (see also Fig.~\ref{lightcurves}). We first searched for
frequencies in all filters of the individual seasons, using SCARGLE
(Scargle 1982) and Phase Dispersion Minimisation (PDM, Stellingwerf
1978) analysis, and the least-squares power spectrum method
(Van\'{\i}\v{c}ek 1971).  The runs in 2005 at SNO and SPMO were
too short (a time span of 2 and 4 days, respectively) to resolve
$\gamma$ Dor type periods. In the other seasons, except in the SNO
subset obtained in November 2007--January 2008, we detected a
frequency near 0.24\,\cd, which is present in all filters. In the SNO
datasets a frequency of 1.73\,\cd\, is also detected. 

 As all individual seasons have different zero points due to
instrumental effects, we aligned the data before merging them. The
alignment was performed in a delicate and iterative process. We
calculated the preliminary frequency solution of the individual
subsets (i.e. we performed a least-squares fit with the frequency
0.24\,\cd), and used the constant of the fit as a measure to re-align
the subsets at the same mean brightness level. In a next step we
determined an improved frequency solution from the merged dataset, and
calculated from the individual subsets the least-squares fit with this
new solution to obtain an improved value of the constant, which was
then used to construct an improved merged dataset. We continued this
iterative process until an optimized multi-periodic solution was found
(see below). The value of the constant of the least-squares fit at the
end of the iterative process did not differ more than 0.2\,mmag from
the constant calculated in the first iteration step, in all filters.
We are aware that this artificial alignment can remove or modify
long-period variations, if present.  To quantify the artificial
shifts, we give a list, for each filter, of the relative correction
between the subsets of the SPMO data and SNO data separately, as well
as the relative offset between SPMO and SNO datasets. The relative
corrections between the SPMO subsets are of the order of 18.1\,mmag ($u$ filter),
21.9\,mmag ($v$ filter), 24.7\,mmag ($b$ filter), and 6.8\,mmag
($y$ filter). For the SNO subsets the corrections are 20.3\,mmag ($u$
filter), 9.7\,mmag ($v$ filter), 6.3\,mmag ($b$ filter), and
6.7\,mmag ($y$ filter). The offset between the SPMO and SNO subsets
are of the order of 20.0\,mmag ($u$ filter), 21.0\,mmag ($v$ filter), 41.0\,mmag
($b$ filter), and 9.0\,mmag ($y$ filter).

In the aligned datasets the frequency near 1.7348\,\cd\, appears besides
0.2342\,\cd, and, moreover has the highest amplitude. More
  low-amplitude frequencies seem to be present in the light variations
  (near 2.538\,\cd\, and 2.253\,\cd), but without further evidence we
  only accept the two frequencies that satisfy the SN$>$4 criterion
  (Breger et al. 1993; Kuschnig et al. 1997): 1.73480(3)\,\cd\, and
  0.23427(5)\,\cd. The SN-level was computed as the average amplitude
  over a frequency interval with a width of 5~\cd\, in an oversampled
  SCARGLE periodogram obtained after final prewhitening.   We
    carefully checked if the two frequencies are independent (their
    sum is close to, but significantly different from, 2\,\cd), and
    found convincing evidence that they indeed both independently
    contribute to the variability. Different stages in the
    prewhitening procedure resulted in the detection of both
    frequencies. Moreover, the same results are obtained when applying the
    Van\'{\i}\v{c}ek method (Van\'{\i}\v{c}ek 1971), which does not rely on the concept of
    prewhitening, since the amplitudes and phases of the known terms
    are recalculated for each new trial frequency.  The frequency
    errors listed in Table~\ref{harmonicfitphot} are calculated using
    the formula described by Montgomery\& O'Donoghue (1999).

\begin{figure}
\centering
\resizebox{1.0\linewidth}{!}{\includegraphics{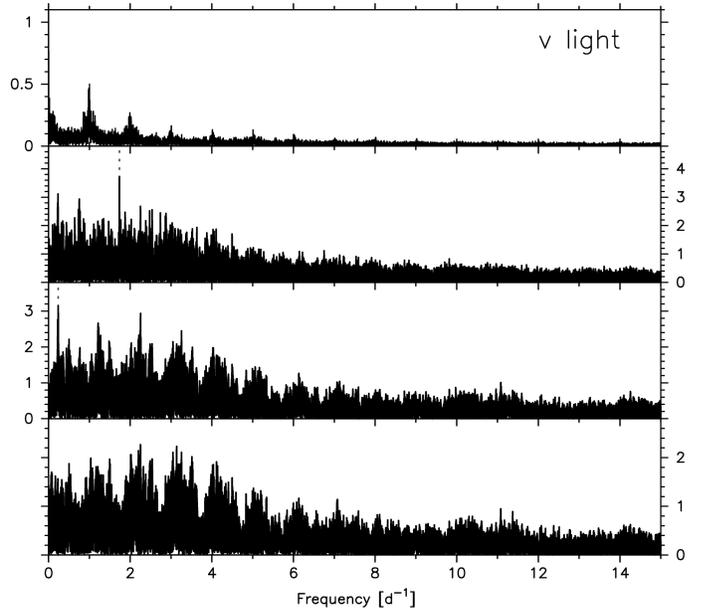}}
\caption{SCARGLE periodograms calculated from the Str\"omgren $v$ data (SPMO and SNO)) of HD\,49434. The top panel gives the associated spectral window, followed by the periodogram of the observed light variations, and the ones calculated after a subsequent removal of 1.73480\,\cd\, and 0.23427\,\cd. These two frequencies are indicated by dashed gray lines. The amplitudes are expressed in millimag.}
\label{scaVlight}
\end{figure}

 We also derived merged datasets (V light) consisting of Str\"omgren
  $y$ data (SPMO and SNO), Johnson V data (KON) and/or HIPPARCOS data.
  For adding the KON and/or HIPPARCOS data we used the same alignment
  procedure as listed above. We note that the individual KON Johnson
  dataset was too short to detect accurate frequencies.
  Unfortunately, an analysis of the HIPPARCOS dataset only ($\Delta T
  = 1095$d; 104 datapoints) resulted in the non-detection of
  frequencies. The frequency search was hampered by a combination of a
  bad observing window (the highest frequency in the observing window
  is 0.2311\,\cd, which is exactly of the order of the
  $\gamma$\,Dor-type frequencies we are looking for) and the
  low-amplitudes of the intrinsic frequencies of HD\,49434, which got
  lost in the noise. In the merged V light datasets we did not find
  other frequencies than the ones detected in the Str\"omgren datasets
  alone. Given the inconvenient time sampling of the HIPPARCOS data,
  and the small sample of KON data, we decide to be conservative and,
  despite the larger time-span, to maintain the frequency accuracy
  derived from the Str\"omgren dataset.  The SCARGLE periodograms of
  the Str\"omgren $v$ light can be found in Fig.~\ref{scaVlight}.  The
  solution of the least-squares fit of this bi-periodic model is given
  in Table~\ref{harmonicfitphot}.  

It is striking to discover that we
  cannot explain the major part of the observed light intensity
  variations of HD\,49434 in our ground-based photometric
  dataset. Several other low-amplitude oscillations probably also contribute to
  the light variability.
From the nightly variations (see bottom right panel in
Fig.~\ref{lightcurves}) we suspect the presence of periods of the
order of a few hours as well. As can be seen at the bottom panel of
Fig.~\ref{scaVlight}, the power excess at this spectral region is low.
To give a quantitative comparison, we computed the mean level of the
noise in the residual dataset for three separate regions: 0--5 \cd,
5--10 \cd\, and 10--15 \cd. We obtain a mean noise level of 0.45 mmag,
0.25 mmag and 0.22 mmag, respectively. We refer to Sect.~\ref{summary}
for a further discussion on the subject.

\begin{table*}
\caption{Amplitudes (in mmag) and phases (in rad) of the frequencies
   $1.73480$~\cd\, and $0.23427$~\cd\, calculated from the {\sl uvby}
   time-series  of HD\,49434 by
   means of a bi-periodic least-squares fit. Errors in units of the last digit are given in parentheses. For each filter the total variance
   reduction of the bi-periodic model and the r.m.s. of the residuals
   is given. Phase = 0 corresponds to $T_0= 2454057.2640$.}
\label{harmonicfitphot}
\begin{tabular} {l cc cc cc cc}
\hline
 \hline
& \multicolumn{2}{c}{Str\"omgren $u$} & \multicolumn{2}{c}{Str\"omgren $v$} & \multicolumn{2}{c}{Str\"omgren $b$} & \multicolumn{2}{c}{Str\"omgren $y$} \\ \hline
   & ampl & phase & ampl & phase & ampl & phase  & ampl & phase  \\ \hline
1.73480\,\cd & 2.5(3) & 5.8(1) & 3.4(2) & 5.84(5) & 3.1(2) & 5.86(5) & 2.4(1) & 5.89(6)  \\
0.23427\,\cd & 1.8(3) & 3.5(2) & 2.5(2) & 3.80(7) & 2.1(2) & 3.80(8) & 1.6(1) & 3.80(8)  \\
\noalign{\smallskip}
\hline
Var. Red. & \multicolumn{2}{c}{5.9\%}  & \multicolumn{2}{c}{24.7\%}& \multicolumn{2}{c}{24.4\%}& \multicolumn{2}{c}{21.2\%} \\
Residual r.m.s. [mmag] & \multicolumn{2}{c}{9.0}  & \multicolumn{2}{c}{6.0}& \multicolumn{2}{c}{6.0}& \multicolumn{2}{c}{5.0} \\
\noalign{\smallskip}
\hline
\end{tabular}
\end{table*}

\subsection{Spectroscopic analysis}
\label{spectroscopy}
\begin{figure}
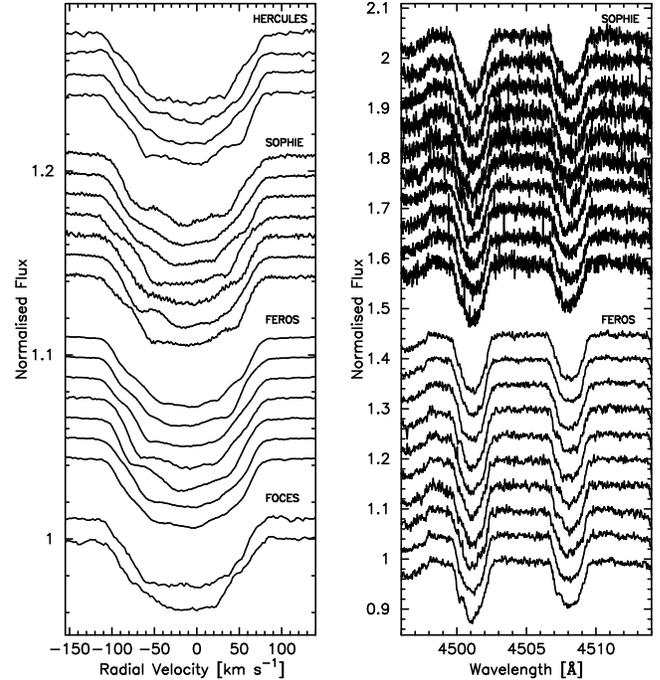

\centering
\begin{tabular}{cc}
\resizebox{0.45\linewidth}{!}{\includegraphics{Uytterhoevenfig4.ps}} &
\resizebox{0.45\linewidth}{!}{\includegraphics{Uytterhoevenfig5.ps}}\\
\end{tabular}
\caption{Left: A selection of LSD profiles calculated from the HERCULES, SOPHIE, FEROS and FOCES spectra of HD\,49434. Right: A selection of SOPHIE and FEROS spectra of HD\,49434, centered on the \TiII\ 4501.273 \AA\ and \FeII\ 4508.288 \AA\ profiles. The spectra are offset for clarity.}
\label{lsdprofiles}
\end{figure}

In addition to individual line profiles, in particular the \TiII\
4501.273 \AA\ and \FeII\ 4508.288 \AA\ profiles (right panel
Fig.~\ref{lsdprofiles}), we also considered deconvoluted profiles
computed through the Least-Squares Deconvolution (LSD) method (Donati
et al. \cite{d97},\cite{d99}).  The LSD method extracts an 'average'
line profile from several hundreds of individual lines by comparing
them with synthetic line masks. We used the spectral line list from
the VALD database (Piskunov et al.\ 1995, Ryabchikova et al.\ 1999,
Kupka et al.\ 1999). We chose a mask with $T_{\rm eff}=7250$ K and
$\log g = 4.0$ and included all elements except He and H. We
calculated the LSD profiles from the regions 4380--4814 \AA\ and
  4960--5550 \AA, which resulted in combining the information of more
than 3100 lines for the FEROS, SOPHIE and FOCES spectra, and 1700
lines for the HERCULES spectra. Through this method we increased the
relatively low SN-ratio of the spectra by creating for each spectrum a
single profile with an average SN-ratio of 5000, 2800, 1900 and 1400
for the FEROS, SOPHIE, HERCULES, and FOCES spectra, respectively.
 As the created 'average' line profiles do not have a
  continuumlevel at 1.0, the LSD profiles were subsequently
  normalised. It turned out that the LSD profiles calculated from
  different instruments have different depths, while the depth of the
  LSD profiles from the same instrument are comparable.  In order to
  homogenize the profiles observed with different instruments, a
  scale-factor was computed from the average LSD profile per dataset,
  and subsequently applied, to rescale all the profiles to comparable
  line depths. We note here that the absolute RV values calculated
  from the LSD profiles do not have a physical meaning, but the
  relative RV changes are intrinsic to the star.  A selection of the
resulting LSD profiles is presented in Fig.~\ref{lsdprofiles}.  The RV
step of the individual LSD profiles were 1 km s$^{-1}$ (SOPHIE), 2 km
s$^{-1}$ (FEROS and FOCES) and 2.5 km s$^{-1}$ (HERCULES). We did not find evidence for a possible binary nature of HD\,49434 in
the line profiles.  We
  discovered the existence of systematic instrumental differences
  between the FEROS, SOPHIE, HERCULES and FOCES spectra. For instance,
  the FOCES spectra seem to be blue shifted over $\sim 1.5$ km
  s$^{-1}$ with respect to the profiles obtained with the FEROS and
  SOPHIE instruments, while the HERCULES spectra are slightly shifted
  towards the red. To correct for the
  instrumental effects, we shifted the spectra from each dataset with
  a fixed RV value ('aligned profiles'). The values were determined as
  the average of the first normalised moment, and were -12.8327\,\kms,
  -13.3156\,\kms, -13.7354\,\kms\, and -15.1830\,\kms\, for HERCULES,
  SOPHIE, FEROS and FOCES spectra, respectively.

We computed the first zero point position of the Fourier transform of
the mean LSD profile of each instrument, which is an indication for
the \vsini\, of the star (Gray 2005), assuming that rotation is the
chief contributor to the line broadening. We obtained the same value
\vsini$ = 87 \pm 1$ km s$^{-1}$ from all FEROS, SOPHIE, HERCULES and
FOCES spectra, within error bars.

 We searched for intrinsic frequencies in two
line-diagnostics: the pixel-to-pixel variations across the line
profile and the moment variations.

\subsubsection{Variations across the line profile}
\label{LPV}
To find variable signals in each pixel along its time variation we
used the two-dimensional Fourier transform analysis method (Intensity
Period Search (IPS) Method, Telting \& Schrijvers 1997).  To optimise
the frequencies and to remove false peaks from the power spectra, we
applied the CLEANEST algorithm (Foster 1995) at the positions of the
first and second zeros of the Fourier transform of the individual
profiles.

\begin{figure}
\centering
\includegraphics[width=8cm]{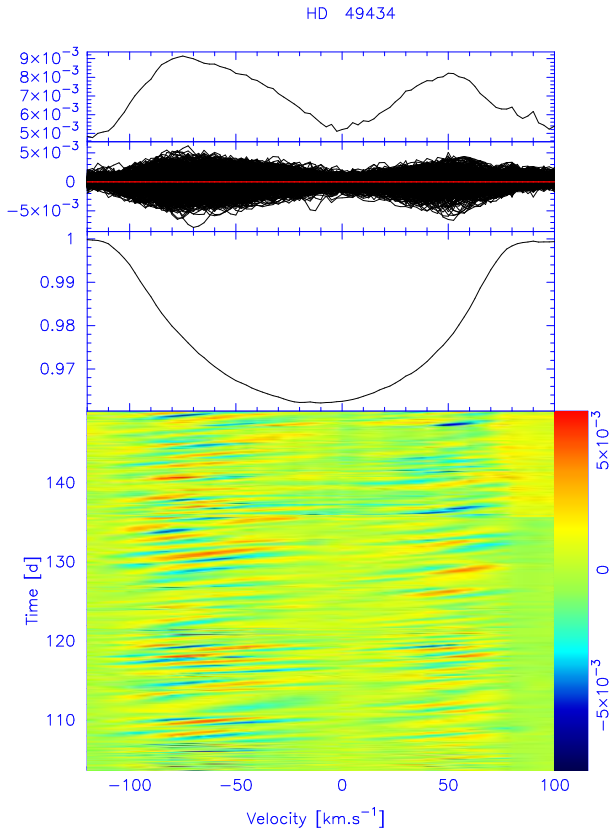}
\caption{Lower part: Bi-dimensional plot of the temporal evolution
(on an arbitrary scale) of the
residual FEROS, SOPHIE and HERCULES LSD profiles of the star  HD\,49434.
A pattern of high frequencies clearly dominates the LPV.
Upper part: from bottom to top are successively represented the mean
spectrum, the individual residual spectra and the dispersion $\sigma$ around
the mean residual. It can be noted that the
variations in the blue wing have higher amplitudes than the ones in the
red wing.}
\label{fig1}
\end{figure}

Figure\,\ref{fig1} presents the two-dimensional dynamic spectra
evolution of the FEROS, SOPHIE and HERCULES LSD profiles. Travelling
bumps are easily seen moving from blue to red. It attracts the
attention that the variations in the blue wing have higher amplitudes
than the ones in the red wing.  This fact has already been reported on
individual line profiles (Mathias et al. \cite{m04}), and is probably
caused by equivalent width (EW) variations of the intrinsic profile
due to local temperature variations on the stellar surface, which
are on their turn resulting from the pulsations (Schrijvers \& Telting
1999). Indeed, they showed that non-adiabatic modes can cause a
non-adiabatic phase-lag in the temperature (or EW) response, which is
observable as an asymmetry in the IPS (see also Fig.~\ref{mode-id}).

We analysed all individual datasets, as well as the combined sets.
Unfortunately, the combination of the datasets resulted in loss of RV
resolution as we had to rebin the profiles to the same RV step. The total time
span of the joined FEROS, SOPHIE and HERCULES spectra (648 spectra)
measures 45.4 days.  Adding the 47 FOCES spectra as well we obtain a
time base of 75 days.

The frequency search across the LSD line profiles (region from  -105 to
101 km s$^{-1}$ in the aligned profiles) resulted in the discovery of at least six frequencies
(see Table~\ref{frequencies}). We accepted only frequencies that
clearly were present in the individual datasets, as well as
frequencies for which we found additional evidence in the moment
variations (see Sect.\,\ref{secmoment}). The dominant frequency is
9.3070(3)\,\cd, followed by 5.3311(3)\,\cd, 12.0332(3) (or
11.0332)\,\cd, 10.1527(7) (or 9.1527)\,\cd, 6.6841(7) (or 7.6841) and
1.4831(8)\,\cd. For three of the frequencies we cannot decide between
the frequency and one of the aliases at $\pm 1$\,\cd. The errors on
the frequencies as calculated from Montgomery \& O'Donoghue (1999) are
given in parentheses. The residual signal after removing the six
frequencies reveals the presence of other periodicities, but given the
low-amplitudes and the fact that we cannot decide between a frequency
and several of its aliases we take a conservative attitude.  Candidate
frequencies that we do not accept without additional evidence are
3.182\,\cd\, and 3.720\,\cd, or one of their aliases.

\begin{figure}
\centering
\resizebox{1.0\linewidth}{!}{\rotatebox{-90}{\includegraphics{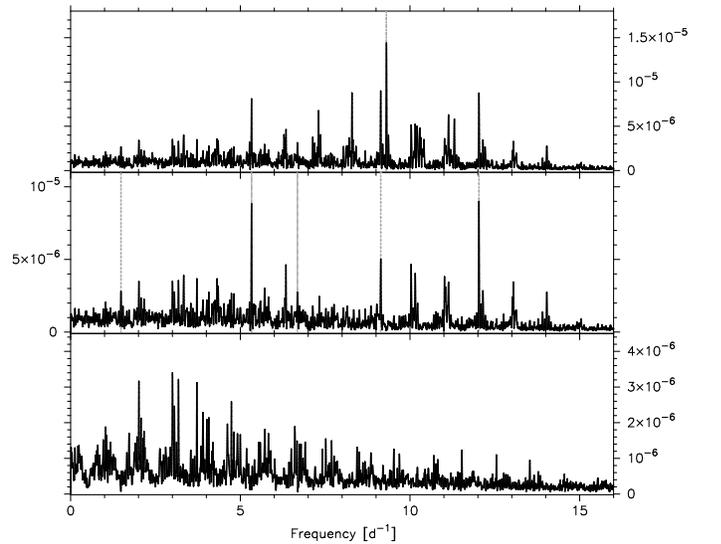}}}
\caption{Summed power over each pixel of the LSD profiles of the
Fourier power spectra calculated from the combined dataset of FEROS,
SOPHIE and HERCULES LSD spectra, which is a measure for the
amplitude of the variability. The top panel gives the power spectra
calculated from the original profiles. The middle panel shows the
power spectra after removal of the dominant frequency (9.302\,\cd).
At the bottom the residual power spectra are shown after prewhitening
the six frequencies 9.3070\,\cd, 5.3311\,\cd, 12.0332\,\cd,
9.1527\,\cd, 6.6841\,\cd\ and 1.483\,\cd, which are indicated by the
light grey dotted lines.}
\label{1dclean}
\end{figure}

 The averaged Fourier power spectrum for the combined dataset of
FEROS, SOPHIE and HERCULES LSD profiles is given in
Fig.~\ref{1dclean}. The bottom two panels give the Fourier spectra
after prewhitening with the dominant first frequency (9.3070\,\cd) and
the residuals after prewhitening the six frequencies described above,
which are indicated by dashed gray lines. 

 Similar and hence consistent frequencies are obtained from the
 pixel-to-pixel analysis of the individual \TiII\ 4501.273 \AA\ and
 \FeII\ 4508.288 \AA\ profiles.

\subsubsection{Moment variations}
\label{secmoment}
\begin{figure}
\centering
\resizebox{1.0\linewidth}{!}{\rotatebox{-90}{\includegraphics{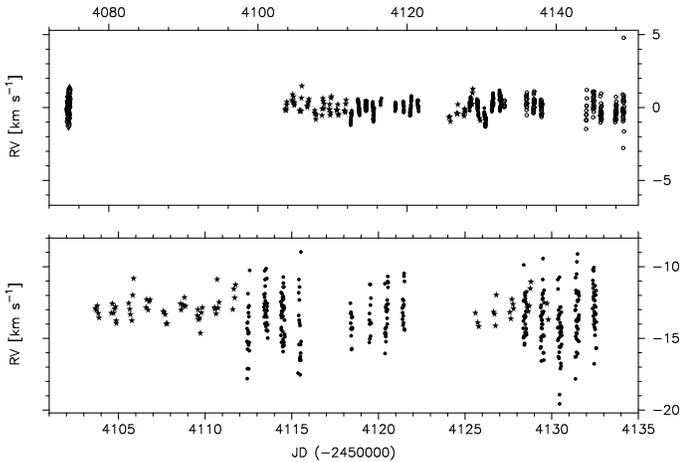}}}
\caption{First normalised moments (RVs) calculated from the FEROS
(stars), SOPHIE (bullets), HERCULES (open circles) and FOCES
(triangles) spectra of HD\,49434. Top panel: $\langle v\rangle$
calculated from the LSD profiles after correction for the small
velocity shifts between different datasets resulting from systematical
instrumental differences. Bottom panel: average value of $\langle
v\rangle$ calculated from the \TiII\ and \FeII\ profiles near 4500
\AA\ of the FEROS and SOPHIE spectra. We used the same  relative scale in y-axis
in top and bottom panel to illustrate the scatter difference between
the RVs measured from the LSD profiles and from the individual
profiles.}
\label{moments}
\end{figure}

A moment analysis was performed considering the EW and the first three
normalised moments ($\langle v\rangle, \langle v^2\rangle, \langle
v^3\rangle$, e.g.\ Aerts et al.\ 1992), calculated from the \TiII\
4501.273 \AA, \FeII\ 4508.288 \AA\ and LSD line profiles, using fixed
integration boundaries (integration domains were $[4498.30,4503.40]$
\AA, $[4505.40,4510.70]$ \AA, and  $[-94.4,91.0]$ \kms,
respectively). We analysed the resulting time-series using the
least-squares power spectrum method and SCARGLE and PDM analyses.
Figure \ref{moments} shows the first normalised moments calculated
from the LSD profiles (upper panel) and from the \TiII\ and \FeII\
profiles (bottom panel).  It is clear that due to the low SN-ratio of
the individual spectra, the different instrumental properties and the
strong dependence of the moment measurements on the SN value and the
normalisation accuracy, the scatter of the moments calculated from the
\TiII\ and \FeII\ profiles is very high, and very different for
individual instruments (FEROS and SOPHIE measurements are represented
by stars and bullets, respectively). A moment analysis of the
individual profiles did not give satisfactory results. Therefore we
describe only the results obtained from the much more accurate LSD
profiles.

\begin{figure}
\centering
\resizebox{1.0\linewidth}{!}{\rotatebox{-90}{\includegraphics{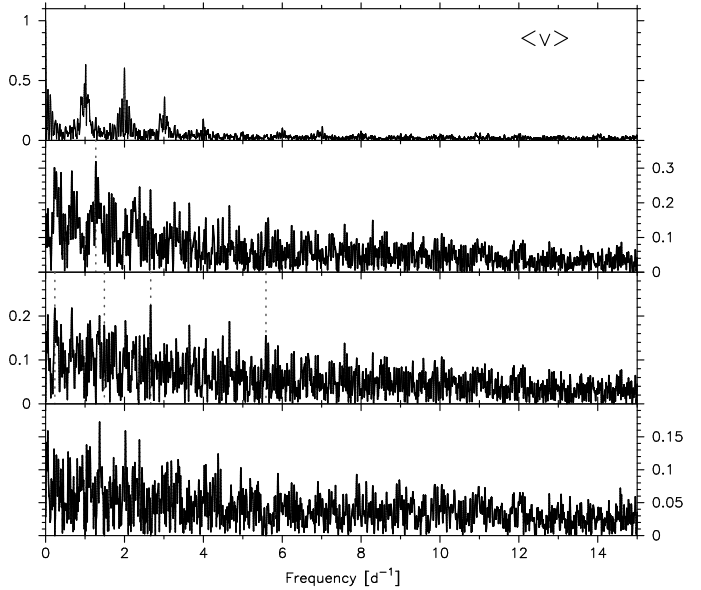}}} 
\caption{SCARGLE periodograms calculated from the first moment of the
combined FEROS, SOPHIE and HERCULES LSD profiles. The top panel gives
the associated spectral window. Subsequently are plotted, from top to
bottom: the periodogram of the first moment, the one after prewhitening with 1.2732
\cd, and the one after additional removal of 2.666\,\cd, 0.234\,\cd,
1.489\,\cd\, and 5.583\,\cd. The listed frequencies are indicated by
light grey lines. The amplitudes are expressed in \kms.}
\label{scarglemom}
\end{figure}

Again we considered the individual as well as the combined data
sets. We could not detect a clear periodicity in the EW variations.
The individual FEROS, HERCULES and FOCES moments present a strong
noise, with no dominant signals.  We present an overview of the
frequencies found in the moments of the combined FEROS, SOPHIE and
HERCULES data, which is illustrated by Table~\ref{freqmodel},
Table~\ref{frequencies} and Fig.~\ref{scarglemom}.  Adding the
one night of lower quality FOCES data did not improve the frequency solution and as they merely added noise to the dataset we 
omitted them in the analysis. We accepted frequencies that satisfied
the SN$>$4 criterion, as well as low-amplitude frequencies for which
we found evidence in more than one line-diagnostics and/or in the
multi-colour variations.

The dominant frequency in all moments is 1.2732(8)\,\cd.  In the first
moment we additionally find 2.666(2)\,\cd, 0.234(1)\,\cd, 1.489(2)\,\cd\,
and 5.58(1)\,\cd. The latter two frequencies have low amplitudes, but
we accept them as they are confirmed in the variations of the second
and higher order ($\langle v^4\rangle, \langle v^5\rangle, \langle
v^6\rangle$) moments and/or by the line-profile analysis.  
On the other hand, the frequency 2.666\,\cd\, is only detected in the
variations of the first moment, and not in any other
diagnostics. However, the first harmonic of this frequency fairly
equals 5.3311\,\cd, which is found in the line-profile analysis. We
note that the contribution of 5.3311\,\cd\, to the variations of the
first moment is not significant.  The frequency 0.234 \cd\, is also
present in multi-colour variations (see Sect.~\ref{multi-colour}).  In
the residuals of the first moment indications for several other
periodicities (e.g.\, 1.369\,\cd\, and 2.024\,\cd) occur, but we need
confirmation from a more extended and accurate dataset, such as CoRoT,
before accepting them as real.  Figure~\ref{scarglemom} shows the
SCARGLE periodogram calculated from the first moment of the combined
dataset.

Besides the dominant frequency 1.2732(8)\,\cd, the second moment shows
variations with 1.489(1)\,\cd\, and 5.583(1)\,\cd. The frequency 1.489\,\cd\ was also detected in the line-profile analysis (see Sect.~\ref{LPV}). We currently accept only these three frequencies, even though the residual time-series seems to hide  other periodicities (e.g.\, 4.629\,\cd).

In the third moment we detect only one additional frequency, 9.308(2)
\cd, which is the dominant frequency found in the pixel-to-pixel
variations across the line-profile.

In summary, in the moment variations we find evidence for a total of
five frequencies, of which two were already detected in the
line-profile variations (see Sect.~\ref{LPV}) and one was discovered
in the photometric dataset (see
Sect.~\ref{multi-colour}). Additionally, we detected the frequency
2.666\,\cd\, in the variations of the first moment, which seems to be
the double period of frequency 5.3311\,\cd. A possible explanation for
the detection of both frequencies is a rotational effect caused by the
presence of a surface inhomogeneity, with the rotational frequency
$f_{\rm rot}=2.666$\,\cd. Given a stellar radius of $R = 1.601
R_{\odot}$ (Masana, Jordi \& Ribas 2006) and a \vsini\, value of
87\,\kms (see Sect.~\ref{spectroscopy}), we find the lower limit
$f_{\rm rot} > 1.07$ \cd\, for HD\,49434. These values of $R$ and
\vsini, and assuming $f_{\rm rot}=2.666$\,\cd, imply an inclination
angle $i = 24^{\circ}$ and an equatorial velocity $v_{\rm eq} = 236$
\kms. These results would make HD\,49434 a fast rotator. Obviously, the
status of the 2.666\,\cd\, frequency as the rotational frequency needs
further investigation.

The solution of the least-squares fit of the model with six
frequencies  is given in Table~\ref{freqmodel}. This
model accounts for 37\%, 30\% and 28\% of the total observed
variability in the first, second and third moment, respectively. We
expect that HD\,49434 oscillates in other additional modes with
amplitudes below our current detection limit.  The continuous and
accurate CoRoT lightcurves might shed a light on this matter.

\begin{table}
\caption[]{Amplitudes (in km s$^{-1}$, (km s$^{-1}$)$^2$ or (km s$^{-1}$)$^3$ for $\langle v\rangle$, $\langle v^2\rangle$ and  $\langle v^3\rangle$, respectively) of the contribution of the frequencies to the variability of the first three moments calculated from the LSD profiles of the combined FEROS, SOPHIE and HERCULES  spectra of  HD\,49434.  The total variance of reduction is given at the bottom.}
\begin{center}
\begin{tabular}{c|ccc} \hline \hline
& $\langle v\rangle$ & $\langle v^2\rangle$ &  $\langle v^3\rangle$ \\ \hline
frequency [\cd] & ampl $\pm$ s.e. & ampl $\pm$ s.e. & ampl $\pm$ s.e.  \\ \hline 
$0.234  $ & $ 0.22\pm0.03 $ & $ 5\pm1 $ & $ 651\pm189  $ \\ 
$1.2732 $ & $ 0.32\pm0.03 $ & $ 16\pm1 $ & $ 2374\pm188 $ \\
$1.489  $ & $ 0.20\pm0.03 $ & $ 12\pm1 $ & $ 1055\pm189  $ \\ 
$5.583  $ & $ 0.14\pm0.03 $ & $ 12\pm1 $ & $ 433\pm185 $ \\ 
$9.308  $ & $ 0.09\pm0.02 $ & $ 3\pm1  $ & $ 1076\pm180   $ \\ 
$2.666  $ & $ 0.17\pm0.03 $ & $ 3\pm1 $ & $ 954\pm190  $ \\ 
 \hline
  frac. var.     &   36.8 \%     & 30.0 \% &  28.4.0 \%      \\ \hline
\end{tabular}
\end{center}
\label{freqmodel}
\end{table}

\subsection{Frequency summary} \label{summary}
Table~\ref{frequencies} gives an overview of all the detected
frequencies together with an indication about the diagnostic (moments,
line profiles or light variations) in which the variation is
prominently present. In total we find nine 'bona fide' frequencies, of
which three show an ambiguity concerning their value. The frequency
2.666(2)\,\cd\, can possibly be interpreted as the rotational
frequency $f_{\rm rot}$ of the star. The frequency 5.3311\,\cd\ might be the first harmonic of the rotational frequency, and might point towards the presence of surface inhomogeneities. This hypothesis needs confirmation. In addition, we propose
other candidate frequencies 3.182\,\cd\, and 3.720\,\cd\, or one of
their aliases, for which we need additional evidence. We did not find
obvious relations (combination effects or aliases) between the
proposed frequencies. In particular, the frequencies at 1.27\,\cd\,
and at 1.73\,\cd (their sum gives 3\,\cd) seem to be really
independent, since the photometric (spectroscopic) data are not well
fitted by the 1.27\,\cd (1.73\,\cd) term.

It immediately attracts
the attention that we detected excess power in two separate regions of
the frequency spectrum: one centered on 1.5\,\cd\, and one at higher
frequencies, between 5 and 12 \cd. The first region is characteristic
for $\gamma$\,Dor pulsators and the second range is typical for
$\delta$\,Sct stars.  The presence of both $p$- and $g$-types
pulsation is not completely unexpected, given the location of the star
very close to the hot border of the $\gamma$\,Dor instability strip,
and within the $\delta$\,Sct one (see Fig.~\ref{CMD}).  We therefore
propose HD\,49434 as a new hybrid $\gamma$\,Dor/$\delta$\,Sct
variable.  In the multi-colour variations we detected only
$\gamma$\,Dor type of variations at significantly high amplitudes.
We barely find traces of the spectroscopic frequencies in the
photometric data (amplitudes $<$ 0.6\,mmag), probably because the
high-degree modes (see Sect.\ref{mode-ID}) are canceled out in the
integrated light.  It is puzzling that the dominating frequency of the
moments 1.2732\,\cd\, is not detected in the pixel-to-pixel variations.
As our frequency model only explains a small portion of
the total observed variability, which is a common problem in
$\gamma$\,Dor star studies (Mathias et al. 2004), we expect that
 other pulsation frequencies are present, but could not be
detected due to the limitations of our ground-based dataset, such as
non-continuous time sampling and a restricted amplitude accuracy. The
exploitation of the continuous CoRoT space data of HD\,49434 promises
a break-through in the understanding of this puzzling, challenging and
very interesting target.

\begin{table}
\caption[]{Overview of the detected frequencies, expressed in\,\cd, in the variability of HD\,49434. The errors on the frequencies, as calculated from Montgomery \& O'Donoghue (1999), are given between brackets. Three frequencies show an ambiguity concerning their value. With a  cross (x) is indicated if the frequency is detected in photometry, in the moments, and/or in the pixel-to-pixel (LPV) variability. It seems that HD\,49434 exhibits both long and short periodic variations and hence can be regarded as a hybrid $\gamma$\,Dor/$\delta$\,Sct star. The frequency 2.666\,\cd\, is possibly the rotational frequenct $f_{\rm rot}$, and 5.3311\,\cd\, its first harmonic.}
\begin{center}
\begin{tabular}{l|ccc|c|c}
\hline \hline
Freq (\cd)  &  $\langle v\rangle$ & $\langle v^2\rangle$ &  $\langle v^3\rangle$ & LPV & photometry \\ \hline
0.23427(5)       &         x            &                      &                       &     &    x       \\
1.2732(8)       &        x            &      x               &      x                &     &            \\
1.4831(8)       &        x            &      x               &                       &  x  &            \\
1.73480(3)       &                     &                      &                       &     &    x       \\
2.666(2)  & x            &                      &                       &     &         \\
5.3311(3)       &                     &                      &                       &  x  &            \\
5.583(1)       &        x            &      x               &                       &     &            \\
9.3070(3)       &                     &                      &      x                &  x  &            \\  
6.6841/7.6841       &                     &                      &                       &  x  &            \\
10.1527/9.1527      &                     &                      &                       &  x  &            \\
12.0332/11.0332      &                     &                      &                       &  x  &            \\ \hline
\end{tabular}
\end{center}
\label{frequencies}
\end{table}

\section{Mode-identification}
\label{mode-ID}
\subsection{Moment Method}
We attempted an identification of the dominant frequency in the
 velocity moments, 1.2732\,\cd, using the Moment Method by Briquet \&
 Aerts (2003). This method scans the discrete $(\ell,m)$ space and
  other continuous velocity parameters, such as the inclination
 angle between the rotation axis and the line of sight and the
 intrinsic width of the line profile, in search for the combination
 that leads to the best fit between the observed and the theoretical
 oscillation amplitude of the first moment. The discriminant value
 $\Sigma$ is an indicator of the goodness of fit: the lower $\Sigma$,
 the better the agreement between theoretical and observed moment
 values. The Moment Method works well for the identification of a
 single mode if its amplitude is much larger than the other detected
 signals. The identification of several modes simultaneously is more
 challenging and requires a very accurate description of the
 frequencies and the amplitudes. As our target HD\,49434 shows several
 low-amplitude frequencies, we focused only on the identification of
 the highest amplitude mode, which is also the only mode detected in
 all three moments.  We calculated $\Sigma$ assuming a stellar mass
 and radius of $M=1.55\,M_{\odot}$ and $R=1.6\,R_{\odot}$, resulting
 in a ratio of horizontal to vertical pulsation amplitude $K=17.403$ for 1.2732\,\cd. As HD\,49434 is a fairly fast rotator
 ($v_{\rm eq}\geq 87$ \kms), we used the formalisms with and without taking
 into account rotational effects (see Townsend 1997).  The resulting
 $\Sigma$ values did not favour a specific solution and did not
 discriminate much between the different $(\ell,m)$ combinations.  The
 results rule out a zonal mode and indicate a $5 \leq \ell \leq 7$
 mode, which is probably tesseral.  We tried to improve the results by
 suppressing the contribution of the additional low-amplitude
 variations by averaging the moments in phase bins of 0.025 of the
 oscillation cycle. Such an approach has been successfully applied to
 other stars (e.g. $\kappa$ Sco, Uytterhoeven et
 al. 2004). Unfortunately this did not lead to satisfactory results
 for HD\,49434. Note that 1.2732\,\cd\, only accounts for 21\% of the
 total variability in the first moment. Reducing the scatter on the
 mono-periodic fit, we calculated $\Sigma$ for a perfect theoretical
 time series with the same amplitude and phase properties as observed,
 which resulted in similar identification results as described
 above. Therefore we conclude that 1.2732\,\cd\ is probably a $5 \leq
 \ell \leq 7$ tesseral mode.

\subsection{Intensity Period Search Method}
\begin{figure}
\centering
\resizebox{1.0\linewidth}{!}{\includegraphics{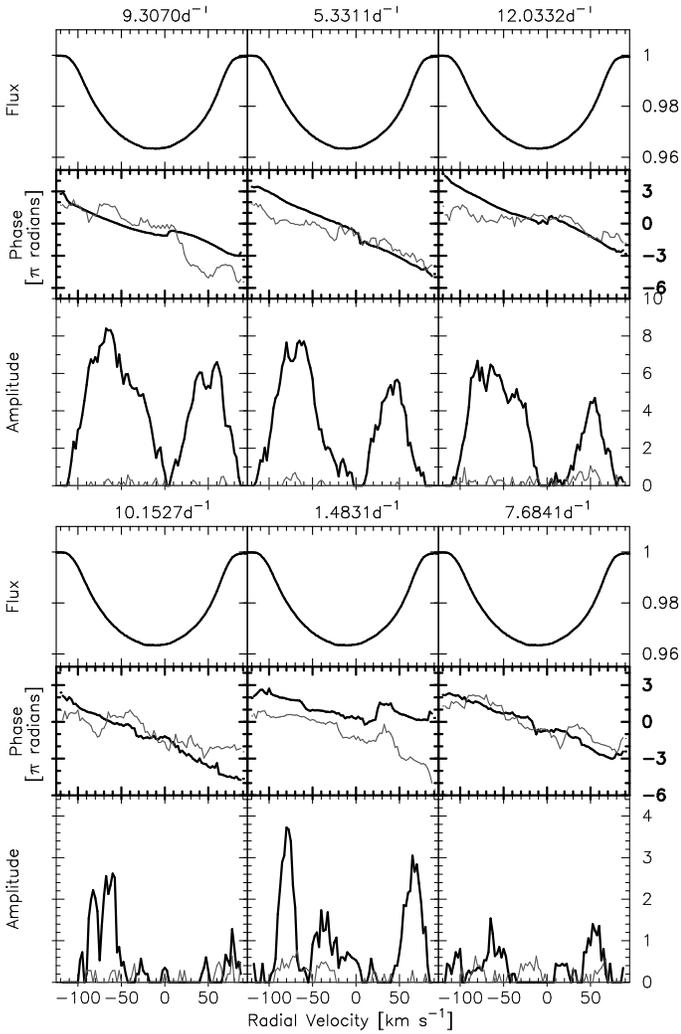}}
\caption{IPS diagrams calculated from the FEROS, SOPHIE and HERCULES
  LSD profiles for the frequencies 1.4831\,\cd, 5.3311\,\cd, 9.3070\,\cd,
  7.6841 (6.6841)\,\cd, 10.1527 (9.1527)\,\cd, and 12.0332 (11.0332)\,\cd. From top
  to bottom: average line profile, phase and amplitude distribution
  across the line profile.  Black and gray lines indicate phase and
  amplitude of a frequency and its harmonic, respectively.}
\label{mode-id}
\end{figure}

The modes detected in the pixel-to-pixel variations of HD\,49434 can
be identified using information on their phase distribution and on the
phase distribution of their first harmonic across the line profile
(IPS Method, Telting \& Schrijvers 1997 ).  From the relations between
the quantum numbers $\ell$ and $m$, and the phase differences with regard to the frequency and its first harmonic,  $\Delta \Psi_{f}$ and  $\Delta \Psi_{2f}$, given by $\ell \approx 0.10 +
1.09|\Delta \Psi_{f}|/\pi$ and $|m| \approx -1.33 + 0.54|\Delta
\Psi_{2f}|/\pi$, as derived by Telting \& Schrijvers (1997), we find
that all modes have a high degree ($3 \leq \ell \leq 8$).  The formal
errors on $\ell$ and $m$ are $\pm 1$ and $\pm 2$, respectively.  For
the detailed results we refer to Table~\ref{IPStable}. We note that we attempted to identify the frequency 5.3311\,\cd\ as well, even though this frequency might be interpretable in terms of surface homogeneities in combination with rotational effects. As this hypothesis needs confirmation, we explore also  an explanation in terms of a non-radial pulsation mode. Figure~\ref{mode-id} shows the mean profile of the observed data
(top), the phase distribution (middle) and amplitude distribution
(bottom) across the line profile for each of the six frequencies
detected in the pixel-to-pixel variations across the line profiles,
listed in Table~\ref{frequencies}. As can be seen in
Figure~\ref{mode-id}, the amplitudes of most  modes are low.
  We note that the candidate frequencies 3.182 \cd\ and 3.720\,\cd (or one of
their aliases), which we did not accept without additional evidence,
show well defined phase distributions, and can be associated to modes
with $\ell \in [2,4]$ and $\ell \in [6,8]$, respectively.

\begin{table}
\caption{Blue-to-red phase differences of the frequencies 1.4831\,\cd, 5.3311\,\cd, 
9.3070\,\cd, 7.6841$\pm$1\,\cd, 10.1527$\pm$1 \c, and  12.0332$\pm$1\,\cd, and their first harmonic, together with estimates for $\ell$ and $m$ according to the relations described by Telting \& Schrijvers (1997), derived from the IPS Method (see Fig.~\ref{mode-id}). The phase differences $\Delta \Psi$ are given in $\pi$ radians.  The identification of 1.2732\,\cd\ by means of the Moment Method is given at the bottom of the table. The frequencies are expressed in\,\cd.}
\begin{center}
\begin{tabular}{ccccc} \hline \hline
 frequency & $\Delta \Psi_f$ & $\Delta \Psi_{2f}$ & $\ell \in$ & $|m| \in$  \\ \hline$1.4831$ & $5.0$ & $6.0$ & $[4,6] $ & $[0,4] $ \\
$5.3311$ & $7.0$ & $6.5 $ & $[6,8] $ & $[0,4] $ \\
$9.3070$ & $5.0$ & $7.0$ & $[4,6]$ & $[0,4] $ \\ \hline
$7.6841 (6.6841)$ & $3.5$ & $7.5$ & $[3,5] $ & $[0,4] $ \\
$10.1527 (9.1527)$& $6.5$ & $\sim6$ & $[6,8] $ & $[0,4] $ \\
$12.0332 (11.0332)$& $6.5$ & $3.5 $ & $[6,8] $ & $[0,2] $ \\ \hline \hline
 frequency &  &  & $\ell \in$ & $|m| $  \\ \hline
$1.2732$  &   &  & $[5,7]$  & $0 < |m| \leq l$ \\ \hline
\end{tabular}
\end{center}
\label{IPStable}
\end{table}

We postpone the identification of the two photometric modes by means
of the amplitude and frequency ratio methods to a subsequent paper, as
the theoretical description of the pulsational character of HD\,49434
is delicate, given its fast rotation ($v_{\rm eq} \geq 87$ \kms). 

\section{Abundance analysis}
We performed an abundance analysis on the average normalised FEROS
spectrum, due to its relatively high mean SN-value compared to the
other data sets.  The synthetic spectra we used, were computed with
the SYNTHE code (Kurucz 1993) ported under GNU Linux by Sbordone et
al. (2005). All the atmospheric models were computed with the
line-blanketed LTE ATLAS9 code, which handles line opacity with the
opacity distribution function method (ODF).  We adopted the atomic
line list from the VALD database ({\it Vienna Atomic Line Database},
Kupka et al. 1999; Ryabchikova et al. 1999).

For the analysis of the data an efficient spectral synthesis method
based on the least-squares optimalisation algorithm was used (Takeda
1995, Bevington 1969). This method allows for the simultaneous
determination of various parameters involved with stellar spectra and
consists in minimisation of the deviation between the theoretical flux
distribution and the observed one.  The synthetic spectrum depends on
stellar parameters such as effective temperature $T_{\rm eff}$,
surface gravity $\log g$, microturbulence $\xi$, rotational
velocity \vsini, radial velocity, and relative abundances of the
elements $\epsilon_i$, where $i$ denotes the individual element.  The
first three parameters were not determined during the iteration
process but were considered as  input values. All the other
above-mentioned parameters can be determined simultaneously because
they produce detectable and different spectral signatures. The
theoretical spectrum was fitted to the average normalised FEROS spectrum.  
 All spectra were normalised using MIDAS procedures. In some parts of the blue wavelength range the continuum level was corrected by comparison between theoretical and observed spectra.

\begin{figure}
\centering
\resizebox{1.0\linewidth}{!}{\includegraphics{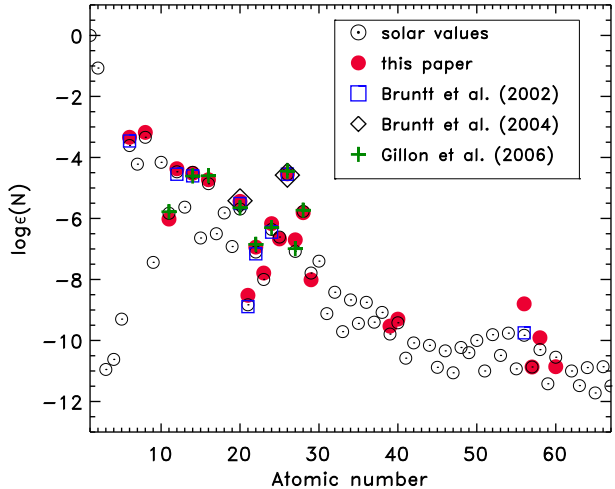}}
\caption{Element abundances of HD\,49434 derived in this paper (red
bullets) compared to the solar values (Grevesse \& Sauval 1998, $\odot$), and
the values obtained for HD\,49434 by Bruntt et al (2002, blue
squares), Bruntt et al (2004, black diamonds) and Gillon \& Magain
(2006, green crosses).}
\label{abundancefig}
\end{figure}

We fixed the input $T_{\rm eff}$ and $\log g$ values to $7300$~K and
$4.2$~dex, respectively (see Sect.~\ref{intro}).
The abundances and rotational velocity resulting from
our analysis are given in Table~\ref{abundance}.  In all calculations
we assumed a microturbulence velocity $\xi$ equal to $2$~km~s$^{-1}$
(Bruntt et al. 2002). The obtained abundances are compared with the
previous results in Fig.~\ref{abundancefig}. As we can see, the
determined chemical composition of HD\,49434 is close to the solar
one. We derived a metallicity of $Z = 0.019 \pm 0.002$ for HD~49434,
assuming solar values from Grevesse \& Sauval (1998) for all elements
not considered in our work. The most important discrepancy was found
for Ba, but for this element only three lines were considered. Our
results are consistent with all values obtained in previous
analyses.  The influence of changes in $T_{\rm eff}$, $\log g$ and microturbulence (e.g.\, Bruntt et al 2002) on
the derived abundances are shown in Table~\ref{abundanceonline},
available in the on-line version of the paper. We derived the
abundances of elements for $T_{\rm eff} = 7300$~K changed by $\pm
400$~K, $\log g= 4.2 \pm 0.1$~dex, and  $\xi=2\pm 1$\kms. The influence of effective
temperature on the abundance is more important, especially for Mg, V
and La, for which the difference is equal to about 0.4~dex. For the
other elements, changes in temperature cause differences in obtained
abundances of about 0.2~dex or lower.

\begin{table}
\caption{Average LTE abundances, $\log \epsilon ({\rm N})$, and standard deviations, $\sigma$ ($\log \epsilon$[H] = 12),  of HD\,49434 determined for $T_{\rm eff}=7300$~K,
$\log g=4.2$\,dex and microturbulence $\xi=2$~km s$^{-1}$.
Standard deviations are given only if more than two spectral parts were used
to determine the element abundance.
In the last two columns the standard solar composition of Grevesse \& Sauval
(1998; Sun 1D) and solar abundances
derived from 3D hydrodynamical models (Asplund et al. 2005; Sun 3D) are
given.}
\label{abundance}
\begin{tabular}{l|cccc}\hline
\hline
N &$\log \epsilon(N)$& $\sigma$ & Sun 1D& Sun 3D  \\
\hline
C & 8.70 &    0.14  &8.52  &8.39\\
O & 8.86 &    $--$  &8.83  &8.66\\
Na& 6.02 &    $--$  &6.33  &6.17\\
Mg& 7.67 &    0.09  &7.58  &7.53\\
Si& 7.52 &    0.47  &7.55  &7.51\\
S & 7.31 &    $--$  &7.33  &7.14\\
Ca& 6.60 &    0.16  &6.36  &6.31\\
Sc& 3.52 &    0.28  &3.17  &3.05\\
Ti& 5.10 &    0.22  &5.02  &4.90\\
V & 4.25 &    0.30  &4.00  &4.00\\
Cr& 5.87 &    0.23  &5.67  &5.64\\
Mn& 5.37 &    0.35  &5.39  &5.39\\
Fe& 7.50 &    0.12  &7.50  &7.45\\
Co& 5.34 &    $--$  &4.92  &4.92\\
Ni& 6.23 &    0.16  &6.25  &6.23\\
Cu& 4.03 &    $--$  &4.21  &4.21\\
Y & 2.51 &    0.18  &2.24  &2.21\\
Zr& 2.74 &    0.15  &2.60  &2.59\\
Ba& 3.24 &    0.06  &2.13  &2.17\\
La& 1.17 &    $--$  &1.17  &1.13\\
Ce& 2.13 &    $--$  &1.58  &1.58\\
Nd& 1.18 &    $--$  &1.50  &1.45\\
\hline
\end{tabular}
\end{table}

\onltab{9}{
\begin{table*}
\caption{Average LTE abundances, $\log \epsilon(N)$, of HD\,49434 with standard deviations $\sigma $ and rotational velocities \vsini\, ([km s$^{-1}$]), determined for the indicated values of $T_{\rm eff}$ ([K]), $\log g$ ([dex]) and
$\xi$ ([km s$^{-1}$]). Standard deviations are given only if more than two spectral parts were used to determine the element abundance.}
\label{abundanceonline}
\begin{tabular}{l|cc|cc|cc|cc|cc|cc|cc} 
\hline
\hline
$T_{\rm eff}$      & \multicolumn{2}{c|}{7300}       & \multicolumn{2}{c|}{7700}       & \multicolumn{2}{c|}{6900}      & \multicolumn{2}{c|}{7300}      &  \multicolumn{2}{c|}{7300} & \multicolumn{2}{c|}{7300} & \multicolumn{2}{c}{7300}
\\
$\log g$               & \multicolumn{2}{c|}{4.20}       & \multicolumn{2}{c|}{4.20}       & \multicolumn{2}{c|}{4.20}      & \multicolumn{2}{c|}{4.10}      &  \multicolumn{2}{c|}{4.30} & \multicolumn{2}{c|}{4.20}  & \multicolumn{2}{c}{4.20} 
\\
$\xi$   &  \multicolumn{2}{c|}{2}         &  \multicolumn{2}{c|}{2}         &  \multicolumn{2}{c|}{2}        &  \multicolumn{2}{c|}{2}        &   \multicolumn{2}{c|}{2}
&  \multicolumn{2}{c|}{1} &  \multicolumn{2}{c}{3}\\
\vsini & \multicolumn{2}{c|}{$89 \pm 3$} & \multicolumn{2}{c|}{$89 \pm 3$} & \multicolumn{2}{c|}{$90 \pm 3$}& \multicolumn{2}{c|}{$89 \pm 3$}& \multicolumn{2}{c|}{$89
\pm 3$} & \multicolumn{2}{c|}{$89 \pm 3$} & \multicolumn{2}{c}{$89 \pm 3$}\\
\hline
C   &8.70 &  0.14   & 8.79  &  0.11  & 8.79  & 0.17  & 8.69 &  0.15  & 8.73  & 0.15  & 8.48 & 0.52 & 8.68 & 0.22 \\
O   &8.86  &   --     & 8.65  &   --    & 9.11  &  --    & 8.83 &   --    & 8.89  & --  & 8.84 & -- & 8.85 & --  \\
Na  &6.02 &   --     & 6.21  &   --    & 5.84  &  --    & 6.05 &   --    & 5.97  &--  & 6.35 & -- & 5.76 & -- \\
Mg  &7.67 &  0.09    &8.02  &  0.08  & 7.26  & 0.06  & 7.75 &  0.06  & 7.63  & 0.09 & 7.95 & 0.34 & 7.62 & 0.38 \\
Si  &7.52 &  0.47   & 7.58  &  0.37  & 7.25  & 0.42  & 7.55 &  0.27  & 7.53  & 0.28 & 7.37 & 0.62 & 7.46 & 0.45\\
S   &7.31 &   --     & 7.53  &   --    & 7.74  &  --    & 7.30 &   --    & 7.33  & --  & 7.16 & -- & 7.21 & -- \\
Ca  &6.60 &  0.16   & 6.70  &  0.14  & 6.38  & 0.13  & 6.62 &  0.07  & 6.65  & 0.10 & 6.77 & 0.57 & 6.48 & 0.16 \\
Sc  &3.52 &  0.28   & 3.60  &  0.17  & 3.53  & 0.24  & 3.45 &  0.20  & 3.53  & 0.21 & 4.18 & 0.89 & 3.35 & 0.26\\
Ti  &5.10 &  0.22   & 5.21  &  0.21  & 4.83  & 0.25  & 4.97 &  0.21  & 5.03  & 0.23 & 4.85 & 0.82 & 4.94 & 0.22 \\
V   &4.25 &  0.30   & 4.65  &   --    & 4.45  &  --    & 4.53 &   --    & 4.56  & -- & 4.38 & -- & 4.18 & --  \\
Cr  &5.87 &  0.23   & 6.05  &  0.18  & 5.72  & 0.19  & 5.89 &  0.24  & 5.90  & 0.27 & 5.87 & 0.51 & 5.79 & 0.22 \\
Mn  &5.37 &  0.35   & 5.68  &  0.18  & 5.07  & 0.40  & 5.44 &  0.18  & 5.44  & 0.19 & 5.32 & 0.96 & 5.41 & 0.39\\
Fe  &7.50 &  0.12   & 7.70  &  0.10  & 7.29  & 0.11  & 7.49 &  0.11  & 7.49  & 0.11 & 7.40 & 0.38 & 7.39 & 0.10\\
Co  &5.34 &   --     & 5.53  &   --    &  --    &  --    & 5.33 &   --    & 5.35  & --  & 4.65 & -- & 5.36 & -- \\
Ni  &6.23 &  0.16   & 6.49  &  0.16  & 6.13  & 0.15  & 6.30 &  0.17  & 6.30  & 0.17 & 6.25 & 0.37 & 6.24 & 0.24\\
Cu  &4.03 &   --     & 4.12  &   --    & 3.79  &  --    & 4.03 &   --    & 4.03  & --  & 3.91 & 0.11 & 4.07 & 0.08  \\
Y   &2.51 &  0.18   & 2.68  &  0.17  & 2.32  & 0.25  & 2.49 &  0.18  & 2.54  & 0.19 & 2.03 & 0.31 & 2.43 & 0.38\\
Zr  &2.74 &  0.15   & 2.98  &   --    & 2.79  &  --    & 2.53 &   --    & 2.84  & -- & 2.78 & 2.30 & 0.71  \\
Ba  &3.24 &  0.06   & 3.47  &   --    & 2.89  &  --    & 3.21 &  0.02  & 3.21  & 0.02 & 3.81 & 0.07 & 2.58 & 0.10\\
La  &1.17 &   --     & 1.56  &   --    & 0.93  &  --    & 1.16 &   --    & 1.17  & --  & 0.89 & -- & -- & -- \\
Ce  &2.13 &   --     & 2.30  &   --    & 1.84  &  --    & 2.07 &   --    & 2.20  & --  & 2.08 & -- & 1.73 & -- \\
Nd  &1.18 &   --     & --     &   --    &  --    &  --    & 1.13 &   --    & 1.27  & -- & -- & -- & --&--   \\
\hline
\end{tabular}
\end{table*}}

\section{Discussion and conclusion}
In this paper we presented frequency analysis and preliminary
spectroscopic mode-identification results of HD\,49434,
obtained from the most extensive ground-based dataset 
for a $\gamma$\,Dor star to date.  
Photometric and spectroscopic observations were performed
between 2005 and 2008 in the framework of the CoRoT ground-based
support campaign.  

Our central finding is the discovery of the hybrid nature of the star:
HD\,49434 pulsates simultaneously in $p$- and $g$-modes.
The frequency analysis clearly shows the presence of at least four
$\gamma$\,Dor-type, as well as  six $\delta$\,Sct-type of
frequencies.  Additional frequencies are expected to contribute to the observed
variability, but have amplitudes near or below our current detection
limit.   All modes, for which an identification was possible, seem to be high-degree
modes ($3 \leq \ell \leq 8$), which make future modelling extremely challenging.

Also the fairly fast rotation of HD\,49434 ($v_{\rm eq} \geq 87$ \kms)
will complicate the modelling since the frequencies in the
observers frame are rotationally modified with respect to the ones in
the stellar rest frame. Actually, a high-degree, $m=-4$ mode can
shift the observed frequencies by 10 \cd, which means that a $p$--mode
may appear with a typical $g$--mode frequency ($m<0$) or vice-versa.
We expect that the CoRoT time-series will provide additional
information on the rotational frequency through the effect of
rotational splitting.  In this context, the frequency 2.666(2)\,\cd\,
can possibly be interpreted as the rotational frequency $f_{\rm rot}$
of the star. Its first harmonic, 5.3311\,\cd, is detected in the
line-profile variations, and if the $f_{\rm rot}$ value is confirmed,
this could imply the presence of surface inhomogeneities. From
the asymmetric behaviour of the amplitude distributions in the IPS
diagrams (see Figs.\,\ref{fig1} and \ref{mode-id}) we find indications
for local temperature variations on the stellar surface (Schrijvers \&
Telting 1999), but we could not detect a clear periodicity in the
corresponding EW variations.  A similar behaviour has been noted in
other stars, such as the $\delta$\,Sct star FG\,Vir (Zima et al. 2006)
and the $\beta$ Cep pulsator $\varepsilon$\,Cen (Schrijvers et
al. 2004).

We encountered several puzzling results. For instance, it is not clear
why we failed to detect the dominating frequency from the moment
variations, 1.2732 \cd, in the variations across the line profiles,
and if this frequency can be linked with the photometric frequency
1.73480 \cd.  It is clear that the pulsational behaviour of HD\,49434 is
very complex  and that the fast rotation adds to this. 
The simultaneous presence of several low-amplitude
variations complicates the frequency search, which often results in an
ambiguous determination of the frequency values (see
Table~\ref{frequencies}).  Moreover, some technical aspects of the
analysis add to the confusion, as the delicate process of the alignment of the data.

The detection of the hybrid nature of HD\,49434 
makes it an interesting
asteroseismic target. The asteroseismic importance of hybrid $\gamma$
Dor/$\delta$\,Sct stars is obvious as both the envelope and the deep
interior of the star can be probed through the study of $p$- and
$g$-modes, respectively, and thus giving a unique opportunity to
constrain the macro-physics of the stellar interior.  The amount of
hybrid detections is steadily growing, although still only a handful
of cases is known. The first discovery of its kind was made in
HD209295 (A9/F0V, Handler et al. 2002). However, this primary of an
eccentric binary might be not a 'bona fide' hybrid variable as the
excitation of the $\gamma$\,Dor-type variations is possibly induced by
tidal interaction. To date at least four other hybrid $\gamma$
Dor/$\delta$\,Sct stars have been reported, of which two were
discovered from MOST satellite data: HD~8801 (A7m, Henry \& Fekel
2005), HD114839 (Am, King et al. 2006), BD+18~4914 (F5, Rowe et
al. 2006), and HD~44195 (F0, Uytterhoeven et al. 2008). The
simultaneous presence of $p$- and $g$-modes has also been
theoretically predicted (Dupret et al. 2004), and hence might not be
as uncommon as it currently seems.  Our results on HD\,49434 show that
the $\delta$\,Sct-type modes have high degrees ($3 \leq \ell \leq 8$,
see Table~\ref{IPStable}), and consequently are only detected in
spectroscopic variations (Table~\ref{frequencies}) as high-degree
modes tend to cancel out in the integrated photometric light.  This
observational selection effect can explain why Handler and Shobbrook
(2002) in their systematic survey for $\delta$\,Sct variations in
$\gamma$\,Dor stars, using only photoelectric photometry, failed to
detect hybrid variables other than HD209295. We can even speculate
that actually a large part of all $\gamma$\,Dor variables pulsate both
in $p$- and $g$-modes, but the high-degree $p$-modes need
spectroscopic detection.  This finding would make the $\gamma$\,Dor
pulsating class extremely suited for asteroseismic studies.  

In a subsequent series of papers we will confront the ground-based
data, enlarged with a new spectroscopic time-series obtained in
January/February 2008, with the actual CoRoT space data with the goal
to perform a deeper investigation and to present a modelling of the
star.

The combination of continuous and accurate space photometry, and
ground-based spectroscopy promises to be the best way to study the
frequency spectrum of HD\,49434 in detail. Results from the MOST
satellite show that not all high-degree spectroscopic modes can be
recovered from photometric space data (e.g.\, $\zeta$ Oph, Walker et
al. 2005), and consequently the spectroscopic observations will play
an important part in the unraveling and understanding of the
pulsational behaviour of the star.

As a matter of fact, the CoRoT
satellite mission will provide an excellent opportunity to investigate
the pulsational behaviour of a large amount of $\gamma$\,Dor variables,
as in the forthcoming years continuous time-series of thousands of
stars in the exo-planetary CCD field will become available. Also, with
at least two hybrid stars selected in the asteroseismic field of
CoRoT, the target of this paper HD\,49434 and HD~44195, we have good
prospects of making progress in modelling and understanding the
relation between $p$- and $g$-mode pulsators.

\begin{acknowledgements}
 We  thank 
Peter De Cat, Luciano Mantegazza, Wolfgang Zima and Andrea
Miglio and the anonymous referee  for useful comments on the
first version of the paper. The FEROS data are being obtained as part of the
ESO Large Programme: LP178.D-0361 (PI: E. Poretti).  This work was
supported by the italian ESS project, contract ASI/INAF I/015/07/0,
WP\,03170, by the Hungarian ESA PECS project No 98022 and by the
European Helio- and Asteroseismology Network (HELAS), a major
international collaboration funded by the European Commission's Sixth
Framework Programme.  KU acknowledges financial support from a
\emph{European Community Marie Curie Intra-European Fellowship},
contract number MEIF-CT-2006-024476. PJA acknowledges financial
support from a Ramon y Cajal contract of the Spanish Ministry of
Education and Science

\end{acknowledgements}

{}

\end{document}